\documentclass[aps,prb,twocolumn,showpacs,preprintnumbers,amsmath,amssymb,superscriptaddress]{revtex4}

\usepackage{graphicx}
\usepackage{dcolumn}
\usepackage{bm}

\newcommand{\bra}[1]{\langle #1|}
\newcommand{\ket}[1]{|#1\rangle}

\begin{document}

\title{Effect of intrinsic spin relaxation on the spin-dependent
cotunneling transport through quantum dots}

\author{Ireneusz Weymann}\email{weymann@amu.edu.pl}
\affiliation{Department of Physics, Adam Mickiewicz University,
61-614 Pozna\'n, Poland}
\author{J\'ozef Barna\'s}
\affiliation{Department of Physics, Adam Mickiewicz University,
61-614 Pozna\'n, Poland} \affiliation{Institute of Molecular
Physics, Polish Academy of Sciences, 60-179 Pozna\'n, Poland}

\date{\today}

\begin{abstract}
Spin-polarized transport through quantum dots is analyzed
theoretically in the cotunneling regime. It is shown that the
zero-bias anomaly, found recently in the antiparallel
configuration, can also exist in the case when one electrode is
magnetic while the other one is nonmagnetic. Physical mechanism of
the anomaly is also discussed. It is demonstrated that intrinsic
spin relaxation in the dot has a significant influence on the
zero-bias maximum in the differential conductance -- the anomaly
becomes enhanced by weak spin-flip scattering in the dot and then
disappears in the limit of fast spin relaxation. Apart from this,
inverse tunnel magnetoresistance has been found in the limit of
fast intrinsic spin relaxation in the dot. The diode-like behavior
of transport characteristics in the cotunneling regime has been
found in the case of quantum dots asymmetrically coupled to the
leads. This behavior may be enhanced by spin-flip relaxation
processes.
\end{abstract}

\pacs{72.25.Mk, 73.63.Kv, 85.75.-d, 73.23.Hk}

\maketitle

\section{Introduction}

Electronic transport through nano-scale systems coupled to
ferromagnetic leads reveals new phenomena which arise due to the
interplay of charge and spin degrees of freedom. In the case of
metallic grains or large quantum dots, this leads to
magnetic-dependent Coulomb blockade and Coulomb oscillations
phenomena \cite{barnas98,weymann03,ernult04}, spin accumulation,
negative differential conductance, and inverse tunnel
magnetoresistance (TMR) effect
\cite{shimada02,yakushiji05,weymannDSET06}. Transport through a
quantum dot with a few electrons reveals further features
resulting from discreteness of the dot energy spectrum and from
electron correlations
\cite{kouwenhoven97,kouwenhoven98,deshmukh02,hanson03,zambuhl04}.
When the dot is weakly coupled to the leads, one can distinguish
two different regimes of electronic transport: sequential and
cotunneling ones \cite{averin90,geerligs90,golovach04}. Sequential
transport through a single-level quantum dot coupled to
ferromagnetic leads was studied quite recently for collinear
\cite{bulka00,rudzinski01} and non-collinear
\cite{koenig03,braun04,braun06} configurations of the magnetic
moments of electrodes. Spin polarized transport in the cotunneling
regime has also been addressed
\cite{weymannPRB05BR,pedersen05,weymannEPJB05}. It has been shown
\cite{weymannPRB05BR} that the interplay of various spin dependent
cotunneling processes gives rise to anomalous behavior of the
differential conductance when the leads' magnetic moments are
antiparallel. This, in turn, leads to the corresponding zero-bias
anomaly in the tunnel magnetoresistance. Moreover, the even-odd
electron number parity effect in tunnel magnetoresistance has also
been found \cite{weymannPRB05a}. In both cases intrinsic spin
relaxation in the dot was neglected, and spin in the dot could
relax only due to cotunneling events.

In this paper we consider transport through single-level quantum
dots coupled to ferromagnetic leads in the cotunneling regime. In
particular, we show how the intrinsic spin relaxation in the dot
affects the dot occupation probabilities, spin accumulation,
conductance and the TMR effect. We analyze the cases of quantum
dots with degenerate and non-degenerate energy levels, which are
symmetrically (equal spin polarizations of the leads) and
asymmetrically (unequal spin polarizations of the leads) coupled
to the leads. The latter case is of particular interest as the
corresponding transport characteristics are highly asymmetric with
respect to the bias reversal. We show that this diode-like
behavior may be enhanced by intrinsic spin-flip relaxation
processes in the dot. Moreover, we show that the zero-bias
anomaly, found previously \cite{weymannPRB05BR} in the
antiparallel configuration, can occur in any magnetically
asymmetric system, in particular when one electrode is nonmagnetic
while the other one is ferromagnetic.

The systems considered in this paper may be realized
experimentally in various ways, including self-assembled dots in
ferromagnetic semiconductors \cite{chye02}, ultrasmall metallic
particles \cite{ralph02}, granular structures \cite{zhang05},
carbon nanotubes \cite{tsukagoshi99,zhao02,sahoo05}, and other
molecules \cite{pasupathy04}. In the case of quantum dots coupled
to nonmagnetic leads, differential conductance in the cotunneling
regime has also been measured experimentally \cite{koganPRL04}.

In section 2 we present the model and theoretical formulation of
the problem. Numerical results for symmetric and asymmetric
systems are shown and discussed in sections 3 and 4, respectively.
Final conclusions are given in section 5.

\section{Model and method}

We consider a single-level quantum dot coupled to ferromagnetic
leads and for simplicity restrict our considerations to the case
when magnetic moments of the leads are either parallel or
antiparallel. The system is described by the Hamiltonian $H=H_{\rm
L}+H_{\rm R}+H_{\rm D}+H_{\rm T}$. The first two terms represent
the left and right reservoirs in the noninteracting particle
approximation, $H_{\nu}=\sum_{k\sigma} \varepsilon_{\nu
k\sigma}c^{\dagger}_{\nu k\sigma} c_{\nu k\sigma}$, where
$\varepsilon_{\nu k\sigma}$ denotes energy of an electron with the
wave number $k$ and spin $\sigma$ in the lead $\nu={\rm L,R}$. The
dot Hamiltonian $H_{\rm D}$ involves two terms, $H_{\rm
D}=\sum_{\sigma=\uparrow,\downarrow}
\varepsilon_{\sigma}d^{\dagger}_{\sigma} d_{\sigma}+U
d^{\dagger}_{\uparrow} d_{\uparrow}
d^{\dagger}_{\downarrow}d_{\downarrow}$, where the first term
describes noninteracting electrons in the dot, whereas the second
one takes into account Coulomb interaction with $U$ denoting the
correlation parameter. The tunneling Hamiltonian reads, $H_{\rm
T}=\sum_{\nu=\rm L,R}\sum_{k\sigma}\left( T_{\nu}c^{\dagger}_{\nu
k\sigma} d_{\sigma}+T^{*}_{\nu}d_{\sigma}^{\dagger}c_{\nu
k\sigma}\right)$, where $T_{\nu}$ denotes the corresponding tunnel
matrix elements. Coupling of the dot to external leads can be
described by the parameters $\Gamma_{\nu}^{\sigma}= 2\pi
|T_{\nu}|^2 \rho_{\nu\sigma}$, with $\rho_{\nu\sigma}$ being the
spin-dependent density of states in the lead $\nu$. When defining
the spin polarization of lead $\nu$ as $p_{\nu}=(\Gamma_{\nu}^{+}-
\Gamma_{\nu}^{-})/ (\Gamma_{\nu}^{+}+ \Gamma_{\nu}^{-})$, the
coupling parameters can be written as
$\Gamma_{\nu}^{+(-)}=\Gamma_{\nu}(1\pm p_{\nu})$, with
$\Gamma_{\nu}= (\Gamma_{\nu}^{+} +\Gamma_{\nu}^{-})/2$. Here,
$\Gamma_{\nu}^{+}$ and $\Gamma_{\nu}^{-}$ correspond to
spin-majority and spin-minority electrons, respectively. In the
following we assume $\Gamma_{\rm L}=\Gamma_{\rm R}\equiv\Gamma/2$.
In the weak coupling regime, typical values of the dot-lead
coupling strength $\Gamma$ are of the order of tens of $\mu$eV,
whereas the measurements are usually carried out at temperatures
of the order of tens of mK \cite{koganPRL04}.

In order to analyze electronic transport in the cotunneling regime
we employ the second-order perturbation theory
\cite{averin90,kang97}. The cotunneling rate for electron
transition from the left to right leads is then given by
\begin{equation}\label{Eq:cotunnelingrate}
\gamma_{\rm LR}=\frac{2\pi}{\hbar}
\left|\sum_{v}\frac{\bra{\Phi_L}H_{T}\ket{\Phi_v}\bra{\Phi_v} H_T
\ket{\Phi_R}} {\varepsilon_i-\varepsilon_{v}}\right|^2\delta
(\varepsilon_i-\varepsilon_f),
\end{equation}
where $\varepsilon_i$ and $\varepsilon_f$ denote the energies of
initial and final states, $\ket{\Phi_\nu}$ is the state with an
electron in the lead $\nu$, whereas $\ket{\Phi_v}$ is a virtual
state with $\varepsilon_{v}$ denoting the corresponding energy.
Generally, one can distinguish between single-barrier and
double-barrier cotunneling as well as between cotunneling
processes that leave unchanged magnetic state of the dot and those
which reverse spin of the dot. The non-spin-flip cotunneling
processes are elastic and coherent, whereas the spin-flip ones are
incoherent (and inelastic when the dot level is spin split).
Unlike the double-barrier processes, the single-barrier ones do
not contribute directly to electric current. However, they can
change the occupation probabilities, and this way also the current
flowing through the system.

In numerical calculations we have included all possible
cotunneling events. Furthermore, we have allowed for intrinsic
spin relaxation processes in the dot \cite{rudzinski01}. Such
processes can result, for instance, from spin-orbit interaction in
the dot or coupling of the electron spin to nuclear spins. We will
not consider a particular microscopic mechanism of the intrinsic
spin-flip processes, but simply assume their presence and describe
them by the relevant spin-flip relaxation time $\tau_{\rm sf}$.
The relaxation processes have been then taken into account {\it
via} a relaxation term in the appropriate master equation for the
occupation probabilities,
\begin{equation}
  0=\sum_{\nu,\nu^\prime = {\rm L,R}}\left(-\gamma_{\nu\nu^\prime}
  ^{\sigma\rightarrow\bar{\sigma}} P_{\sigma} +
  \gamma_{\nu\nu^\prime} ^{\bar{\sigma}\rightarrow\sigma}
  P_{\bar{\sigma}}\right)
  -\frac{2}{\tau_{\rm sf}}\frac{P_\sigma e^{\beta\varepsilon_\sigma} -
   P_{\bar{\sigma}} e^{\beta\varepsilon_{\bar{\sigma}}} }
   {e^{\beta\varepsilon_\sigma}+e^{\beta\varepsilon_{\bar{\sigma}}}},
\end{equation}
where $\beta=1/(k_{\rm B}T)$, $P_\sigma$ denotes the probability
that the dot is occupied by a spin-$\sigma$ electron, and
$\gamma_{\nu\nu^\prime} ^{\sigma\rightarrow\bar{\sigma}}$ is the
cotunneling rate from lead $\nu$ to lead $\nu^\prime$ with a
change of the dot spin from $\sigma$ to $\bar{\sigma}$
($\bar{\sigma}=-\sigma$). The last term describes the spin
relaxation processes. In the case of spin-degenerate dot level
this term is reduced to $-(P_\sigma - P_{\bar{\sigma}})/\tau_{\rm
sf}$.

We consider the case when the dot is singly occupied at
equilibrium ($\varepsilon_\sigma<0$, $\varepsilon_\sigma+U>0$) and
the system is in a deep Coulomb blockade regime; $\Gamma,k_{\rm
B}T \ll |\varepsilon_\sigma|,\varepsilon_\sigma+U$. In such a case
the sequential tunneling is exponentially suppressed, and
cotunneling gives the dominant contribution to electric current
\cite{weymannPRB05BR}. Both non-spin-flip and spin-flip
cotunneling processes are then allowed and have to be taken into
account.

In the following we will distinguish between the fast and slow
spin relaxation limits. The former (latter) limit corresponds to
the situation when the time between successive cotunneling events,
$\tau_{\rm cot}$, is significantly  longer (shorter) than the
intrinsic spin relaxation time $\tau_{\rm sf}$. A typical spin
relaxation time for quantum dots can be relatively long, up to
$\mu$s \cite{fujisawa02,elzerman04}. On the other hand, the time
between successive cotunneling events can be estimated taking into
account the fact that the rate of spin-flip cotunneling is
generally larger than that of non-spin-flip cotunneling (for a
finite parameter $U$). Assuming
$\varepsilon_\uparrow=\varepsilon_\downarrow=\varepsilon$, one
then finds
\begin{equation}\label{Eq:tcot}
  \tau_{\rm cot}\approx
  \frac{h\varepsilon^2(\varepsilon+U)^2}{AU^2\Gamma^2}\,,
\end{equation}
with $A={\rm max}\{|eV|,k_{\rm B}T \}$ and $h=2\pi\hbar$. Assuming
typical parameters \cite{koganPRL04}, one can roughly estimate
$\tau_{\rm cot}$ to range from $10^{-3}$ ns to $1$ ns. Thus,
having spin relaxation time in the dot and the system parameters
entering Eq.~(\ref{Eq:tcot}), one can estimate whether the
system's behavior corresponds to the fast or slow spin relaxation
regimes. Both the time between successive cotunneling events and
the spin relaxation time depend on the system parameters and can
be tuned, for example by applying a weak magnetic field, varying
coupling to nuclear spins, etc. Thus, the regimes of fast and slow
spin relaxation are achievable experimentally.

Electric current flowing through the system from the left to right
leads is given by
\begin{equation}
  I = e \sum_{\sigma\sigma'} P_\sigma \left[
  \gamma^{\sigma \rightarrow \sigma'}_{\rm LR} -
  \gamma^{\sigma \rightarrow \sigma'}_{\rm RL} \right].
\end{equation}
In the following we discuss the influence of intrinsic spin
relaxation on differential conductance and tunnel
magnetoresistance of the quantum dot in the cotunneling regime
both in the presence and absence of external magnetic field $B$.
The TMR effect is defined as
\begin{equation}\label{Eq:TMR}
{\rm TMR} = \frac{I_{\rm P}-I_{\rm AP}}{I_{\rm AP}} \,,
\end{equation}
with $I_{\rm P}$ ($I_{\rm AP}$) denoting the current flowing
through the system in the parallel (antiparallel) magnetic
configuration at a constant bias voltage $V$ applied to the
system.

\section{Quantum dots symmetrically coupled to the leads}

First, let us proceed with the case of quantum dots coupled to
ferromagnetic leads with equal spin polarizations $(p_L=p_R\equiv
p)$. Our considerations will also include the cases of quantum
dots with non-degenerate and degenerate energy levels. The former
situation is discussed in the following subsection.

\subsection{The case of spin-degenerate dot level}

\begin{figure}[t]
  \includegraphics[width=0.7\columnwidth]{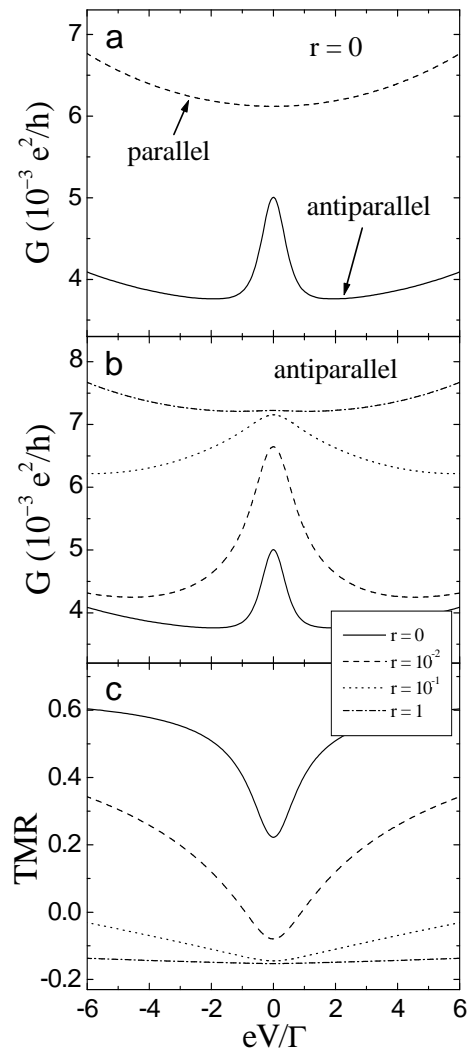}
  \caption{\label{Fig1} The differential conductance in the parallel
  and antiparallel configurations (a,b) and tunnel magnetoresistance
  (c) as a function of the bias voltage for different spin
  relaxation $r=h/(\tau_{\rm sf}\Gamma)$. The parameters are:
  $k_{\rm B}T=0.2\Gamma$,
  $\varepsilon =-15\Gamma$, $U=30\Gamma$,
  and $p_{\rm L}=p_{\rm R}=0.5$.}
\end{figure}

In the case of $B=0$ the dot level is spin degenerate,
$\varepsilon_{\uparrow}= \varepsilon_{\downarrow} =\varepsilon$.
The differential conductance $G=dI/dV$ for parallel and
antiparallel magnetic configurations is shown in Fig.
\ref{Fig1}(a) for the case of no intrinsic spin relaxation in the
dot ($1/\tau_{\rm sf}\rightarrow 0$). In the parallel
configuration one finds typical parabolic behavior of the
conductance with increasing transport voltage. However, this is
not the case for antiparallel configuration, where a maximum in
the differential conductance appears in the small bias voltage
regime. This anomalous behavior of the differential conductance,
in turn, leads to respective minimum in the tunnel
magnetoresistance, as shown by the solid curve in Fig.
\ref{Fig1}(c).

\begin{figure}[t]
  \includegraphics[width=0.7\columnwidth]{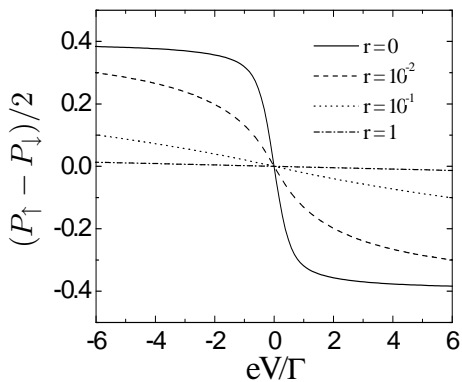}
  \caption{\label{accumulation} The spin accumulation
  in the antiparallel configuration
  as a function of the bias voltage for different spin
  relaxation $r=h/(\tau_{\rm sf}\Gamma)$. The parameters are the
  same as in Fig.~\ref{Fig1}.}
\end{figure}

Physical mechanism of the zero-bias anomaly was considered in
Ref.~[\onlinecite{weymannPRB05BR}] for the case of no intrinsic
spin relaxation. It was pointed out that the anomaly results from
the relative difference between the fastest cotunneling events
(which involve only the majority electrons of the leads)
contributing directly to the charge current in the parallel and
antiparallel configurations. In the parallel configuration, the
main contribution to electric current comes from the non-spin-flip
cotunneling processes, whereas spin-flip cotunneling is dominant
for antiparallel alignment. In addition, the spin asymmetry in
tunneling processes leads to a nonequilibrium spin accumulation in
the dot in the antiparallel configuration, which increases with
increasing bias voltage. The spin accumulation, defined as
$(P_\uparrow - P_\downarrow)/2$, is shown in
Fig.~\ref{accumulation} as a function of the bias voltage for
different spin relaxation times. For the parameters assumed here
and for negative bias ($eV <0$, current flows from right to left,
while electrons flow from left to right), there is a larger
probability to find in the dot a spin-up electron than a spin-down
one. This, in turn, diminishes the contribution coming from the
fastest spin-flip cotunneling processes. However, there are also
single-barrier spin-flip cotunneling events which reverse spin of
the dot and open the system for the fastest transport processes.
The rate of single-barrier processes is proportional to
temperature (but independent of the applied voltage), whereas that
of double-barrier processes increases with the transport voltage,
being roughly independent of temperature. Thus, when $|eV|\lesssim
k_{\rm B}T$, the single-barrier processes play a significant role
-- they decrease the spin accumulation in the dot and open the
system for the fastest cotunneling events. Consequently, the main
contribution to current comes then from the most probable
spin-flip cotunneling processes. When $|eV|\gg k_{\rm B}T$, the
relative role of single-barrier processes is diminished due to an
increased role of double-barrier cotunneling, and the differential
conductance decreases. The zero-bias maximum in differential
conductance appears then as a result of the competition between
single- and double-barrier cotunneling processes.

The anomalous behavior of differential conductance in the
antiparallel configuration results from spin asymmetry of
tunneling processes. This asymmetry leads to nonequivalent
occupation of the dot by spin-up and spin-down electrons (spin
accumulation), as illustrated in Fig.~\ref{accumulation}. From the
above discussion follows that the spin accumulation is crucial for
the occurrence of the zero-bias anomaly. Thus, one may expect that
intrinsic spin-flip processes in the dot should suppress the
anomaly. Indeed, intrinsic spin relaxation in the dot
significantly modifies the conductance maximum at zero bias, and
totally suppresses this anomalous behavior in the limit of fast
spin relaxation. This behavior is displayed in Fig.~\ref{Fig1}(b),
where different curves correspond to different values of the
parameter $r$ defined as $r=h/(\tau_{\rm sf}\Gamma)$. Thus, $r=0$
describes the case with no intrinsic spin relaxation, whereas the
curves corresponding to nonzero $r$ describe the influence of
intrinsic relaxation processes. First of all, one can note that
small amount of intrinsic spin-flip processes enhances the
zero-bias anomaly [see the curve for $r=10^{-2}$ in
Fig.~\ref{Fig1}(b)]. This is because such processes play then a
role similar to that of single-barrier spin-flip cotunneling. In
the case of fast spin relaxation, on the other hand, the spin
accumulation is suppressed and the anomaly disappears, as can be
seen in Fig.~\ref{Fig1}(b) and Fig.~\ref{accumulation} for $r=1$.
It is also worth noting that spin-flip processes in the dot
enhance the overall conductance in the antiparallel configuration.
In the parallel configuration, however, the differential
conductance does not depend on intrinsic relaxation. This is
because there is no spin accumulation in the parallel
configuration, and consequently the intrinsic spin-flip processes
in the dot do not play any role.

\begin{figure}[t]
  \includegraphics[width=0.74\columnwidth]{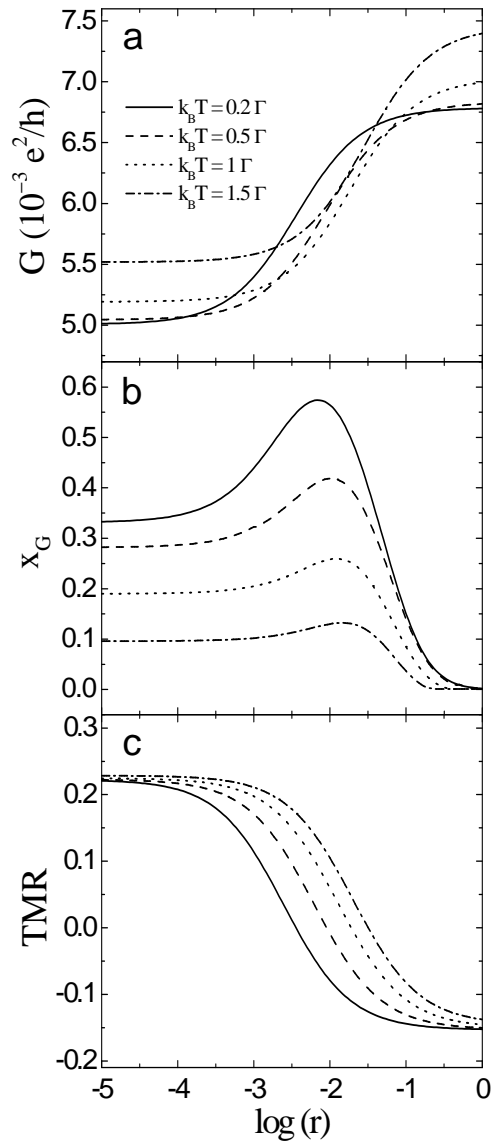}
  \caption{\label{Fig3} The spin relaxation time dependence of the
  linear conductance in the antiparallel configuration (a)
  and the relative height of the zero-bias anomaly
  in differential conductance (b),
  as well as the linear-response TMR (c).
  The parameters are the same as in Fig.~\ref{Fig1}.}
\end{figure}

In the case of a deep Coulomb blockade one can take only the
lowest order corrections in $x/y$, with $x=k_{\rm B}T,|eV|$ and
$y=\vert\varepsilon\vert,\varepsilon +U$, and derive some
approximate formulas for the differential conductance. For the
parallel configuration, the differential conductance can be then
expressed as
\begin{equation}\label{Eq:GP}
  G^{\rm P} = \frac{e^2}{h}\frac{\Gamma^2}{2} \left[
    \frac{1}{\varepsilon^2} + \frac{1}{(\varepsilon +U)^2}
    + \frac{1-p^2}{|\varepsilon| (\varepsilon +U)} \right].
\end{equation}
In the antiparallel configuration and for $|eV| \ll k_{\rm B}T$,
the conductance is given by the expression
\begin{eqnarray}\label{Eq:GAPmax}
  G^{\rm AP}_{\rm max} &=& \frac{e^2}{h}
  \frac{\Gamma^2}{2} \bigg\{(1-p^2) \left[
    \frac{1}{\varepsilon^2} + \frac{1}{(\varepsilon +U)^2}
    + \frac{1}{|\varepsilon| (\varepsilon +U)} \right]\nonumber\\
  &&+\frac{p^2U^2h/\tau_{\rm sf}}{\varepsilon^2(\varepsilon+U)^2h
  /\tau_{\rm sf}+ k_{\rm B}TU^2\Gamma^2}
  \bigg\},
\end{eqnarray}
which describes the maximum of the differential conductance at
zero bias for arbitrary value of $r$. On the other hand, when
$|eV| \gg k_{\rm B}T$ and $r\ll 1$ one finds the formula
\begin{eqnarray}\label{Eq:GAPmin}
  \lefteqn{  G^{\rm AP}_{\rm min} = \frac{e^2}{h}
  \frac{\Gamma^2}{2}\bigg\{ \frac{1-p^2}{1+p^2} \left[
    \frac{1}{\varepsilon^2} + \frac{1}{(\varepsilon +U)^2}
    + \frac{1-p^2}{|\varepsilon| (\varepsilon +U)}
    \right] }\nonumber \\
  && + \frac{4p^2}{1+p^2}  \frac{U^2h/\tau_{\rm sf}}{2\varepsilon^2
  (\varepsilon+U)^2 h/\tau_{\rm sf}+ (1+p^2)k_{\rm B}TU^2\Gamma^2}
  \bigg\},
\end{eqnarray}
which in turn approximates the local minimum value of the
differential conductance. The expression for $G^{\rm AP}_{\rm
min}$ for arbitrary spin relaxation time is too complex to be
presented here. It is clearly evident from the above expressions
that the conductance depends on intrinsic spin relaxation only in
the antiparallel configuration, in agreement with numerical
results. The variation of the linear conductance in the
antiparallel configuration with the parameter $r$ is displayed in
Fig.~\ref{Fig3}(a) for several values of temperature. As one can
see, there is a clear crossover between the limits of fast and
slow spin relaxation in the dot. The crossover depends on
temperature and is shifted towards shorter relaxation times as the
temperature increases.

In order to describe the zero-bias anomaly in the differential
conductance in the antiparallel configuration, it is useful to
introduce the relative height of the maximum at $V=0$, defined as
$x_{\rm G} = (G^{\rm AP}_{\rm max}- G^{\rm AP}_{\rm min})/G^{\rm
AP}_{\rm min}$. The variation of $x_{\rm G}$ with the spin
relaxation time is shown in Fig.~\ref{Fig3}(b) for different
temperatures. First of all, one can see that the relative height
depends on $r$ in a nonmonotonic way. In the limit of slow spin
relaxation and low temperature one finds $x_{\rm G}=4p^2/(3-p^2)$,
whereas in the limit of fast spin relaxation $x_{\rm G}$ tends to
zero, independently of temperature. Indeed, in the limit of fast
spin relaxation $G^{\rm AP}_{\rm max}$ and $G^{\rm AP}_{\rm min}$
are equal and given by
\begin{equation}
  G^{\rm AP} = \frac{e^2}{h}\frac{\Gamma^2}{2} \left[
    \frac{1}{\varepsilon^2} + \frac{1}{(\varepsilon +U)^2}
    + \frac{1+p^2}{|\varepsilon| (\varepsilon +U)} \right].
\end{equation}
On the other hand, the relative height exhibits a maximum for a
certain value of spin relaxation. This maximum, in turn, is
decreased and shifts towards shorter relaxation times with
increasing temperature, see Fig.~\ref{Fig3}(b). One can show that
the maximum occurs when the relaxation time is approximately given
by
\begin{equation}\label{Eq:tmax}
  \tau_{\rm sf}^{\rm max}\approx
  \frac{h\varepsilon^2\left(\varepsilon+U\right)^2}
  {k_{\rm B}T U^2 \Gamma^2} \,.
\end{equation}
When comparing Eqs (\ref{Eq:tcot}) and (\ref{Eq:tmax}) one can
conclude that the maximum in the relative height of the
conductance anomaly occurs when the spin relaxation time
$\tau_{\rm sf}$ is roughly equal to the time between successive
cotunneling events, $\tau_{\rm cot}$, in the zero bias limit. This
also means that the maximum separates the fast and slow spin
relaxation regimes, in agreement with Fig.~\ref{Fig3}. The
enhancement of $x_{\rm G}$ can range from a few percent for small
spin polarization of the leads, $p\lesssim 0.2$, to several
hundred percent for $p\gtrsim 0.9$. For the parameters assumed in
Fig.~\ref{Fig3}(b), the relative height of the zero-bias maximum
in differential conductance for $\tau_{\rm sf}=\tau_{\rm sf}^{\rm
max}$ is enhanced by about $65\%$ as compared to the case of
$\tau_{\rm sf}\rightarrow\infty$. Assuming typical parameters for
the quantum dot \cite{koganPRL04}, one can estimate $\tau_{\rm
sf}^{\rm max}$ to be $\tau_{\rm sf}^{\rm max}\approx 50$ns, which
is accessible experimentally. \cite{fujisawa02,elzerman04}

Intrinsic spin-flip scattering in the dot modifies conductance in
the antiparallel configuration, and thus also the tunnel
magnetoresistance, as shown in Fig.~\ref{Fig1}(c). Since the
zero-bias maximum in conductance is suppressed in the fast spin
relaxation limit, the corresponding dip in TMR at small voltages
is also suppressed by the intrinsic relaxation processes. More
specifically, the dip in TMR broadens with increasing $r$ and
disappears in the limit of fast relaxation (see the curve for
$r=1$).

An interesting and new feature of TMR in the presence of spin-flip
scattering in the dot is the crossover from positive to negative
values when $r$ increases, as illustrated in Fig.~\ref{Fig1}(c).
Thus, the difference between conductances in the parallel and
antiparallel magnetic configurations persists even for fast spin
relaxation in the dot, contrary to the sequential tunneling
regime, where such a difference disappears \cite{rudzinski01}.
This seemingly counterintuitive behavior can be understood by
taking into account the following two facts: (i) absence of spin
accumulation in the dot for fast spin relaxation ($P_\uparrow
=P_\downarrow $), and (ii) difference in the fastest cotunneling
processes contributing to the current in the two magnetic
configurations. As the point (i) is rather obvious in view of the
results presented in Fig.~\ref{accumulation}, the point (ii)
requires additional clarification. The fastest double-barrier
cotunneling processes involve only the majority-spin electrons of
the two leads --  thus, in the parallel configuration the fastest
cotunneling processes are the non-spin-flip ones. They take place
either {\it via} the empty-dot virtual state (for one orientation
of the dot spin) or {\it via} the doubly occupied dot virtual
state (for the second orientation of the dot spin). The dominant
contribution to the current is then proportional to
$1/\varepsilon^2 + 1/(\varepsilon +U)^2$. On the other hand, in
the antiparallel magnetic configuration the fastest cotunneling
processes are the spin-flip ones, which can occur only for one
particular orientation of the dot spin. However, for this spin
orientation cotunneling can take place {\it via} both empty and
doubly occupied dot virtual states. The corresponding dominant
contribution to electric current is then proportional to
$[1/\varepsilon- 1/(\varepsilon +U)]^2=1/\varepsilon^2 +
1/(\varepsilon +U)^2 -2/[\varepsilon (\varepsilon +U)]$. It is
thus clear that the difference in currents flowing through the
system in the antiparallel and parallel configurations is equal to
$-2/[\varepsilon (\varepsilon +U)]$, which results from the
interference term. Since $\varepsilon <0$ and $\varepsilon +U>0$,
this interference contribution is positive. As a result, the
current in the antiparallel configuration is larger than the
current in the parallel configuration.

An analysis similar to that done above for differential
conductance [see Eqs~(\ref{Eq:GP})-(\ref{Eq:GAPmin})] can be
performed for tunnel magnetoresistance. The minimum in TMR at zero
bias in the case of a symmetric Anderson model can be then
expressed as
\begin{equation}
  {\rm TMR_{min} } = \frac{2p^2 \left(4k_{\rm B}T\Gamma^2 -\varepsilon^2
  h/\tau_{\rm sf} \right)}{12(1-p^2)k_{\rm B}T\Gamma^2+(3+p^2)\varepsilon^2
  h/\tau_{\rm sf}},
\end{equation}
whereas for $|eV|\gg k_{\rm B}T$ and $r\ll 1$ one finds
\begin{equation}
  {\rm TMR_{max} } = \frac{2p^2 \left[ 2(3-p^2)k_{\rm B}T\Gamma^2
  -\varepsilon^2 h/\tau_{\rm sf} \right]}
  {2(1-p^2)(3-p^2)k_{\rm B}T\Gamma^2+(3+p^2)\varepsilon^2
  h/\tau_{\rm sf}}.
\end{equation}
The latter formula approximates the value of TMR corresponding to
the bias voltage at which the differential conductance has a local
minimum. In the slow spin relaxation limit one finds ${\rm
TMR_{min} }=2p^2/(3-3p^2)$, and ${\rm TMR_{max} }=2p^2/(1-p^2)$.
However, in the limit of fast spin relaxation TMR becomes negative
and is given by
\begin{equation}
{\rm TMR_{min} }={\rm TMR_{max} }=-2p^2/(3+p^2),
\end{equation}
which is consistent with numerical results displayed in
Fig.~\ref{Fig1}(c).

The explicit variation of the minimum in TMR at zero bias with
spin relaxation time is shown in Fig.~\ref{Fig3}(c). The
transition between slow and fast spin relaxation in the dot is
clearly evident. This transition is associated with a change of
the TMR sign from positive to negative.

\subsection{The case of spin-split dot level}

\begin{figure}[t]
  \includegraphics[width=0.7\columnwidth,height=9.5cm]{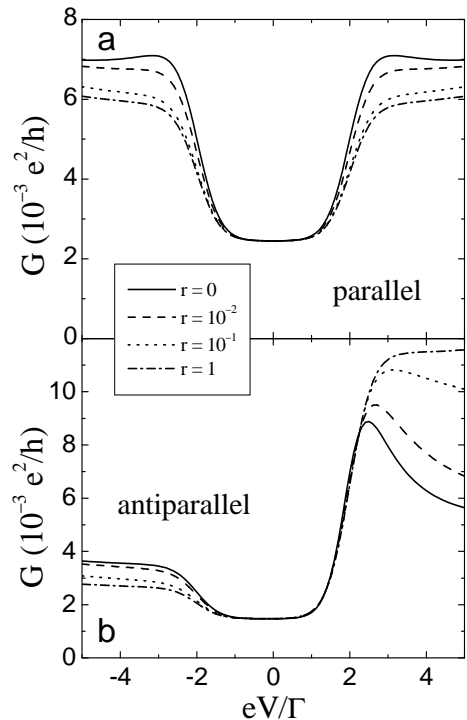}
  \caption{\label{Fig4} The differential conductance in the
  nonlinear response regime for different spin relaxation in the
  parallel (a) and antiparallel (b) configurations. The parameters
  are: $k_{\rm B}T=0.2\Gamma$, $\varepsilon_{\uparrow}=-16\Gamma$,
  $\varepsilon_{\downarrow}=-14\Gamma$, $U=30\Gamma$ and $p=0.5$.}
\end{figure}

The discussion up to now was limited to the case of degenerate dot
level. The situation changes when
$\varepsilon_\uparrow\ne\varepsilon_\downarrow$, e.g., due to an
external magnetic field. The magnetic field splits the dot level,
consequently the occupation probabilities for the spin-up and
spin-down electrons are not equal at equilibrium ($V=0$). The
level splitting is described by the parameter
$\Delta=\varepsilon_\downarrow- \varepsilon_\uparrow$, where the
magnetic field is assumed to be along the magnetic moment of the
left electrode. In Fig.~\ref{Fig4} we show the bias voltage
dependence of the differential conductance in the parallel and
antiparallel configurations for different values of parameter $r$.
In the limit of no intrinsic spin relaxation in the dot (solid
line in Fig.~\ref{Fig4}) and at low bias voltage, the dot is
occupied by a spin-up electron and the current flows mainly due to
non-spin-flip cotunneling. The spin-flip cotunneling processes are
suppressed for $\vert \Delta\vert \gtrsim \vert eV\vert, k_{\rm
B}T$, which results in the steps in differential conductance at
$\vert\Delta\vert \simeq \vert eV\vert$. The suppression of
spin-flip inelastic cotunneling was recently used as a tool to
determine the spectroscopic $g$-factor \cite{koganPRL04}. When
$\vert eV\vert$ becomes larger than $\vert\Delta\vert$, spin-flip
cotunneling is allowed, consequently the conductance increases.
However, there is a large asymmetry of differential conductance in
the antiparallel configuration with respect to the bias reversal.
To understand this asymmetry, it is crucial to realize that when
the splitting $\Delta =\varepsilon_\downarrow -
\varepsilon_\uparrow$ is larger than $k_{\rm B}T$, the
single-barrier spin-flip cotunneling processes can occur only when
the dot is occupied by a spin-down electron. This follows simply
from the energy conservation rule. (The situation may be changed
for negative $\Delta$.) Thus, the single-barrier processes can
assist the fastest double-barrier cotunneling processes, but only
for positive bias. This is because the fastest processes can occur
when the dot is occupied by a spin-down electron for negative bias
and by a spin-up electron for positive bias, leading to larger
conductance for positive than for negative bias voltage. This is
indeed the case in the numerically calculated characteristics
shown in Fig.~\ref{Fig4}(b). No such asymmetry occurs in the
parallel configuration [see Fig.~\ref{Fig4}(a)], as now the system
is fully symmetric with respect to bias reversal.

The situation changes when intrinsic spin-flip relaxation
processes occur in the dot. For the parameters assumed in
Fig.~\ref{Fig4}, the spin relaxation in the dot affects the
conductance only for $|eV|\gtrsim|\Delta|$, while for
$|eV|\lesssim|\Delta|$ the conductance is basically independent of
$r$ (see Fig.~\ref{Fig4}). This is because for
$|eV|\lesssim|\Delta|$ and $|\Delta|\gg k_{\rm B}T$, the dot is
predominantly occupied by a spin-up electron and the transitions
to the spin-down state due to relaxation processes are
energetically forbidden. As a consequence, the current flows
mainly due to non-spin-flip cotunneling, irrespective of spin
relaxation time. This scenario holds for both magnetic
configurations of the system.

When $|eV|\gtrsim|\Delta|$, the spin-flip cotunneling processes
can take place and the dot can be either in the spin-up or
spin-down state. In the parallel configuration,
Fig.~\ref{Fig4}(a), the conductance is slightly reduced by the
spin-flip relaxation processes. This can be understood by
realizing the fact that the fastest non-spin-flip cotunneling
processes in the parallel configuration are more probable when the
dot is occupied by a spin-down electron than by a spin-up one [due
to smaller energy denominator, see
Eq.~(\ref{Eq:cotunnelingrate})]. Since the spin-down state (as
that of larger energy) relaxes relatively fast to the spin-up
state (which has definitely smaller energy), this leads to a
reduction in the conductance. On the other hand, the differential
conductance in the antiparallel configuration is enhanced by the
relaxation processes for positive bias and diminished for negative
bias voltages. Consider first the situation for positive bias. As
already discussed above for $r=0$, an important role in that
transport regime is played by the single-barrier spin-flip
cotunneling processes, which open the system for the fast
double-barrier cotunneling by reversing spin of the dot from the
spin-down to the spin-up state. The relaxation processes play a
role similar to that of the single-barrier cotunneling, and lead
to a certain increase in the conductance. For negative bias
voltage, in turn, the fast double-barrier cotunneling processes
occur when the dot is occupied by a spin-down electron. The
probability of such events is decreased by spin relaxation,
leading to a reduced conductance. An interesting consequence of
the enhancement (reduction) of the differential conductance for
positive (negative) bias voltage is an increase of the asymmetry
with respect to the bias reversal -- see the curve for $r=1$ in
Fig.~\ref{Fig4}(b).

\section{Quantum dots asymmetrically coupled to the leads}

An interesting situation occurs when the quantum dot is coupled
asymmetrically to the left and right leads $(p_{\rm L}\neq p_{\rm
R})$. For this situation we first present the results in the case
of degenerate dot level and then proceed with the discussion of
the case when the degeneracy is lifted, e.g. by an external
magnetic field.

\subsection{The case of spin-degenerate dot level}

\begin{figure}[t]
  \includegraphics[width=0.7\columnwidth]{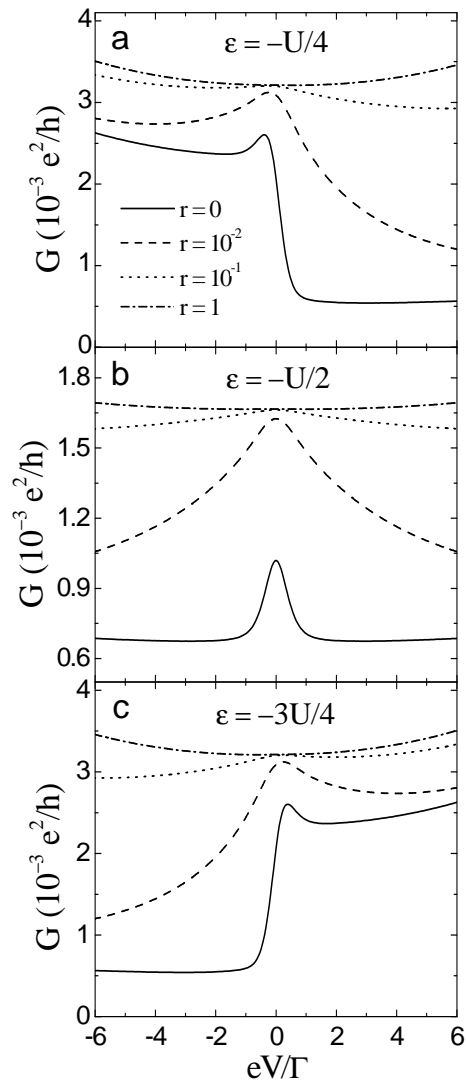}
  \caption{\label{Fig5} The differential conductance in the
  nonlinear response regime for the asymmetric (a,c) and symmetric
  (b) Anderson model for different spin relaxation. The parameters
  are: $k_{\rm B}T=0.2\Gamma$, $U=60\Gamma$, $p_{\rm L}=0.95$ and
  $p_{\rm R}=0$.}
\end{figure}

The differential conductance for a system with one electrode
nonmagnetic and the other one made of a ferromagnet with large
spin polarization (in the following referred to as strong
ferromagnet) is shown in Fig.~\ref{Fig5}. The parts (a) and (c)
correspond to the cases when the dot is described by an asymmetric
Anderson model ($\vert\varepsilon\vert\ne\varepsilon +U$). The
difference between (a) and (c) is due to different position of the
dot level $\varepsilon$, and consequently also $\varepsilon +U$,
with respect to the Fermi level at equilibrium. More precisely,
(a) and (c) are symmetric in the sense that
$\vert\varepsilon\vert$ in (a) is equal to $\varepsilon +U$ in
(c), and vice-versa, $\vert \varepsilon\vert$ in (c) is equal to
$\varepsilon +U$ in (a). Part (b), in turn, corresponds to the dot
described by a symmetric Anderson model, with $\vert
\varepsilon\vert$ equal to $\varepsilon +U$.

Consider first the situation in the absence of intrinsic spin
relaxation in the dot (solid curves in Fig.~\ref{Fig5}). There is
then a significant asymmetry of electric current with respect to
the bias reversal in the situations displayed in
Fig.~\ref{Fig5}(a) and (c), whereas no such an asymmetry occurs
for the symmetric Anderson model [part (b)]. Another feature of
the curves shown in Fig.~\ref{Fig5} is the presence of zero-bias
anomaly in the case described by the symmetric Anderson model
[part (b)]. This anomaly is similar to that shown in
Fig.~\ref{Fig1}(a) for antiparallel configuration. The anomaly
still exists, although it is less pronounced when the dot is
described by the asymmetric Anderson model, as clearly visible in
Fig.~\ref{Fig5}(a) and (c).

Let us discuss now the physical mechanism of the asymmetry with
respect to the bias reversal displayed in Fig.~\ref{Fig5}(a) and
(c). When $\vert eV\vert \gg k_{\rm B}T$, one can neglect the
influence of single-barrier cotunneling \cite{weymannPRB05BR}. The
cotunneling processes which transfer charge from one lead to
another take place {\it via} two possible virtual states -- empty
dot (an electron residing in the dot tunnels to one of the leads
and another electron from the second lead enters the dot) and
doubly occupied dot (an electron of spin opposite to that in the
dot enters the dot and then one of the two electrons leaves the
dot). Consider first the situation shown in Fig.~\ref{Fig5}(a) for
positive bias ($eV>0$, electrons flow from right to left, i.e.,
from normal metal to strong ferromagnet), and assume for clarity
of discussion that the strong ferromagnet is a half-metallic one
with full spin polarization (only spin-up electrons can then
tunnel to the left lead). When a spin-down electron enters the
dot, it has no possibility to leave the dot for a long time. The
allowed cotunneling processes occur then  {\it via } doubly
occupied dot  virtual states. In the absence of intrinsic spin
relaxation in the dot, the only processes which can reverse the
dot spin are the single-barrier cotunneling ones, which however
play a minor role when $k_{\rm B}T\ll \vert eV\vert$. Thus, the
current flows due to non-spin-flip cotunneling {\it via}
doubly-occupied dot virtual states, whereas cotunneling through
empty-dot virtual states is suppressed. The situation is changed
for negative bias (electrons flow from strong ferromagnet to
normal metal). Now, the dot is mostly occupied by a spin-up
electron, which suppresses cotunneling {\it via} doubly-occupied
dot virtual state and the only contribution comes from cotunneling
{\it via} empty-dot virtual state. The ratio of cotunneling rates
through the empty dot and doubly occupied dot virtual states is
approximately equal to $\xi=[\varepsilon/ (\varepsilon+U) ]^{-2}$.
In the situation presented in Fig.~\ref{Fig5}(a) one finds $\xi\gg
1$. Accordingly, the conductance for negative bias is much larger
than for positive bias voltage. In turn, in the case displayed in
Fig.~\ref{Fig5}(c) the ratio of the respective cotunneling rates
is exactly equal to the inverse of this ratio corresponding to the
case shown in Fig.~\ref{Fig5}(a). Thus, now we have $\xi\ll 1$,
and the conductance is smaller for negative bias than for
positive. In turn, in the case presented in Fig.~\ref{Fig5}(b) the
ratio  $\xi$ is equal to 1, consequently both cotunneling rates
are equal and the conductance is symmetric with respect to the
bias reversal.

When $\vert eV \vert$ becomes of the order of $k_{\rm B}T$ or
smaller, the rate of single-barrier cotunneling is of the order of
the rate of double-barrier cotunneling. Therefore, the
single-barrier processes can play an important role in transport.
More precisely, single-barrier cotunneling processes can reverse
spin of an electron in the dot and thus can open the system for
the fast cotunneling processes. This behavior can be observed in
Fig.~\ref{Fig5}(a) and (c). In the symmetric case shown in
Fig.~\ref{Fig5}(b), the single-barrier processes lead to the
zero-bias anomaly similar to that shown in Fig.~\ref{Fig1}(a) for
antiparallel configuration in a system with two ferromagnetic
electrodes coupled symmetrically to the dot. The mechanism of the
anomaly is the same, i.e., the single-barrier cotunneling, which
occurs for $k_{\rm B}T \gg \vert eV\vert$, opens the system for
the fast cotunneling processes. This leads to an increase in
conductance in the small bias range. The zero-bias anomaly exists
also in asymmetric cases illustrated in Fig.~\ref{Fig5}(a) and
(c). However, the maximum in conductance is slightly shifted away
from $V=0$ and is much less pronounced.

Intrinsic spin-flip processes in the dot have similar influence on
electronic transport as in the symmetric case studied in the
previous section. As before, relaxation processes remove the
asymmetry with respect to the bias reversal and suppress the
zero-bias anomaly. Thus, the diode-like behavior can appear only
in the limit of slow spin relaxation, and is suppressed in the
limit of fast spin relaxation, as shown in Fig.~\ref{Fig5} by the
curves corresponding to $r=1$.

Some analytical expressions for the conductance can be obtained in
the limit $\vert eV\vert \gg k_{\rm B}T$ and when assuming $p_{\rm
L}=p$, $p_{\rm R}=0$. The approximate formulas for the
differential conductance then read
\begin{eqnarray}
  G &=& \frac{e^2}{h}\frac{\Gamma^2}{2} \left[
    \frac{1}{(\varepsilon+U)^2} + \frac{(1-p^2)U}
    {\varepsilon^2(\varepsilon+U)} \right.\nonumber\\
    &&\left.+ \frac{p^2\varepsilon^2 U (\varepsilon+U)^3}
    {\left[\varepsilon^2 (\varepsilon+U)^2-
    eVU^2\Gamma^2 \tau_{\rm sf}/(4h)\right]^2}
    \right]
\end{eqnarray}
for positive bias, and
\begin{eqnarray}
  G &=& \frac{e^2}{h}\frac{\Gamma^2}{2} \left[
    \frac{1}{\varepsilon^2} + \frac{(1-p^2)U}
    {\vert\varepsilon\vert(\varepsilon+U)^2} \right.\nonumber\\
    &&\left.+\frac{p^2\vert\varepsilon\vert^3 U (\varepsilon+U)^2}
    {\left[\varepsilon^2 (\varepsilon+U)^2+
    eVU^2\Gamma^2 \tau_{\rm sf}/(4h)\right]^2}
    \right]
\end{eqnarray}
for negative bias. In order to describe the diode operation it is
useful to define the ratio  $\alpha$ of the conductances for
positive and negative bias voltages. In the limit of long spin
relaxation this ratio is given by
\begin{equation}
\alpha = \frac{\varepsilon^2+(1-p^2)(\varepsilon+U)U}
{(\varepsilon+U)^2+(1-p^2)\vert\varepsilon\vert U}.
\end{equation}
It is also worth noting that the operation of the diode can be
tuned by the gate voltage (which shifts energy of the dot level).

\begin{figure}[t]
  \includegraphics[width=0.7\columnwidth]{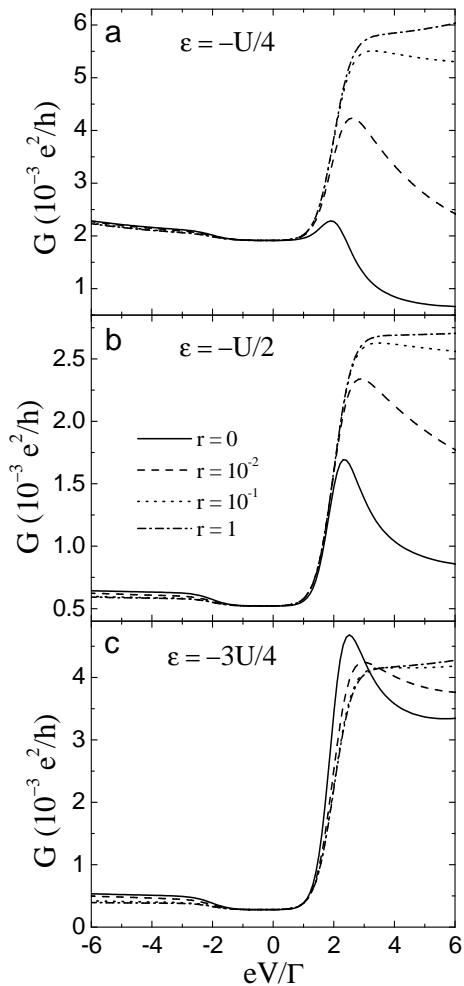}
  \caption{\label{Fig6} The differential conductance in the
  nonlinear response regime for different spin relaxation. The
  parameters are: $k_{\rm B}T=0.2\Gamma$, $\varepsilon_{\uparrow}=
  \varepsilon-\Delta/2$, $\varepsilon_{\downarrow}=
  \varepsilon+\Delta/2$, $\Delta=2\Gamma$, $U=60\Gamma$, $p_{\rm
  L}=0.95$ and $p_{\rm R}=0$.}
\end{figure}

\subsection{The case of spin-split dot level}

In Figure~\ref{Fig6} we illustrate the influence of spin splitting
of the dot level due to an applied magnetic field,
$\varepsilon_\uparrow\neq\varepsilon_\downarrow$. The bias voltage
dependence of the differential conductance is shown there for
different values of the parameter $r$, for symmetric (b) and
asymmetric (a,c) Anderson models. The parameters used in numerical
calculations are the same as in Fig.~\ref{Fig5}, except for the
dot level energy which was spin degenerate in Fig.~\ref{Fig5} and
now is spin split.

Let us first focus on Fig.~\ref{Fig6}(a), which corresponds to the
case of asymmetric Anderson model, $\varepsilon=-U/4$. Two new
features of the conductance in the limit of no intrinsic spin
relaxation appear in this figure. First, the conductance is now
suppressed above a certain finite positive value of the bias
voltage, and not for $eV>0$ as in the case shown in
Fig.~\ref{Fig5}(a). Strictly speaking, the conductance drops when
$eV$ is equal to the level splitting $\Delta$, $eV\approx \Delta$.
The shift of the main slope in the conductance from $eV=0$ to
$eV\approx \Delta$ is due to the fact that the dot is blocked for
cotunneling {\it via} the empty state by a spin-down electron only
when $eV$ exceeds the level splitting. This results from the
energy conservation. Thus, in the case shown in Fig.~\ref{Fig5}(a)
the main drop of the conductance was at $eV=0$, whereas in
Fig.~\ref{Fig6}(a) it is at $eV\approx\Delta$. When a spin-down
electron appears in the dot, then, for $eV<\Delta$, it can always
tunnel back to the source electrode and open the dot for fast
cotunneling through the empty state.

The second feature of the differential conductance in the $r=0$
case is the peak at $eV\approx\Delta$, and a shallow plateau for
$\vert eV\vert \lesssim\Delta$. The origin of the peak is similar
to the origin of the zero-bias anomaly in the case of
spin-degenerate dot level. On the other hand, the plateau in the
small bias range is due to suppression of the spin-flip
cotunneling for $|eV|\lesssim\Delta$, similarly as in
Fig.~\ref{Fig4}. However, the plateau is now very weak and
shallow. This is a consequence of the minor role of the spin-flip
cotunneling, which results from the strong spin polarization of
ferromagnetic electrode.

Qualitatively similar behavior may be also observed for
$\varepsilon=-3U/4$. This situation is shown in Fig.~\ref{Fig6}(c)
and is analogous to the one illustrated in Fig.~\ref{Fig5}(c). In
the case of $\varepsilon=-U/2$ shown in Fig.~\ref{Fig6}(b), the
influence of magnetic field is qualitatively similar to that found
in the case of a quantum dot coupled to two ferromagnetic leads in
the antiparallel configuration, see Fig.~\ref{Fig4}.

Transport characteristics are modified by the spin relaxation in
the dot. As before, the changes occur for $|eV| \gtrsim |\Delta|$.
For positive bias voltage, the spin relaxation processes modify
the conductance mainly for $\varepsilon=-U/4$ and
$\varepsilon=-U/2$ [Fig.~\ref{Fig6}(a,b)], while for
$\varepsilon=-3U/4$ [Fig.~\ref{Fig6}(c)] the changes are
significantly smaller. For negative bias, on the other hand, the
conductance is slightly suppressed by the spin relaxation. This is
because for negative bias the dot is mainly occupied by a spin-up
electron, and spin relaxation from the spin-up to spin-down dot
states is suppressed due to $\Delta \gg k_{\rm B}T$. Contrarily,
for positive bias, $|eV| \gtrsim |\Delta|$, and for $\tau_{\rm
sf}\rightarrow\infty$, the dot is mostly occupied by a spin-down
electron. Spin relaxation processes in the dot lead then to a
certain increase in the occupation probability of the spin-up dot
level ($P_\uparrow\approx 1$ in the long spin relaxation time
limit). This in turn may give rise to an increase in the
differential conductance.

\section{Concluding remarks}

We have considered numerically and analytically the influence of
intrinsic spin relaxation in the dot on the differential
conductance and the TMR effect in the cotunneling regime. It has
been  shown that the zero-bias anomaly in the antiparallel
configuration is enhanced in the weak spin-flip scattering regime
in the dot. The optimal conditions for a maximum enhancement have
also been derived. However, in the limit of fast spin relaxation
both the zero-bias anomaly and the dip in tunnel magnetoresistance
at zero bias become suppressed. Moreover, we have found the
inversion of sign of tunnel magnetoresistance in the presence of
fast intrinsic spin relaxation in the dot.

We have also demonstrated that the zero-bias anomaly exists in
systems with only one ferromagnetic electrode (the second one may
be nonmagnetic). Such devices display diode-like transport
characteristics. The diode-like behavior exists also when the
system is in an external magnetic field, but the operation range
is changed, i.e., the bias voltage at which the conductance drops
depends on the level splitting (magnetic field). Furthermore, we
have shown that this behavior may be enhanced by intrinsic spin
relaxation processes in the dot.

The work was supported by the Polish State Committee for
Scientific Research through the projects PBZ/KBN/044/P03/2001 and
2 P03B 116 25, and by EU through RTN 'Spintronics' (contract
HPRN-CT-2000-000302). We also acknowledge discussions with Jan
Martinek and J\"urgen K\"onig.


\end{document}